\documentclass[a4paper,12pt]{article}

\usepackage{graphicx}
\usepackage{amsmath}
\usepackage{amssymb}

\makeatletter
\renewcommand{\section}{\@startsection{section}{1}{0mm}{\baselineskip}{\baselineskip}{\normalsize\bfseries}}
\renewcommand{\subsection}{\@startsection{subsection}{2}{0mm}{\baselineskip}{\baselineskip}{\normalsize\itshape}}
\makeatother

\title{\bf Reply to comment on `Surface thermodynamics and surface stress for deformable bodies'}

\author{\bf Juan Olives\\
\small CINaM-CNRS\thanks{Associated to Aix-Marseille Universit\'e.}, Campus de Luminy, case 913, 13288 Marseille cedex 9, France\\
\small E-mail: olives@cinam.univ-mrs.fr}

\date{}

\begin{document}

\maketitle

\begin{abstract}

The above comment and a previous letter by the same author reveal a great misunderstanding of what Eulerian and Lagrangian quantities are, and a confusion between the deformation of an element of a surface and the creation of a new element of a surface. Surface thermodynamics is complex because the surface quantities are not `intuitive' (as surface excesses on some dividing surface) and the thermodynamic variables of the state of a surface are {\it a priori} completely unknown. This is why we introduced a new concept (`ideal transformation') and presented detailed proof, leading to the determination of the `local' thermodynamic variables of the state of the surface, the exact expression of the work of deformation of the surface, and the definition of surface stress, for any deformable body (Olives 2010 {\it J. Phys.: Condens. Matter} {\bf 22} 085005). These results are not obvious (despite their similarity with some expressions in {\it volume} thermodynamics). We explicitly write the Eulerian forms of (i) the relation between the surface grand potential per unit area, the surface stress and the surface strain, showing its exact equivalence with the Lagrangian form, and (ii) the variation of the surface energy due to both the deformation of an element of the surface and the creation of a new element of the surface.

\end{abstract}

\maketitle

\noindent The above comment is based on the previous letter \cite{Gutman:1995}. We present here some comments on these papers and about surface thermodynamics and surface stress for deformable bodies.

After (2) of the comment, the author says that our definition of $\gamma_0$ \cite{Olives:2010}
\begin{gather}
\gamma_0 = U_{a_0} - T\,S_{a_0} - \sum_i \mu_i\,m_{i,a_0} \label{defgamma0}
\end{gather}
(excess of grand potential, per unit area in the reference state) only `relates to a fluid material', whereas (21) of \cite{Olives:2010}
\begin{gather}
\delta U_{a_0} = T\,\delta S_{a_0} + \pi_{\rm s} \cdot \delta e_{\rm s}
+ \sum_i \mu_i\,\delta m_{i,a_0} \label{surfaceinternalenergy1}
\end{gather}
only `relates to an elastic body'. This is incorrect. Obviously, at the surface of {\it any deformable body} $\rm b$ (in contact with a fluid $\rm f$), we may define the excesses (on some dividing surface $\rm S_{bf}$, in the present state, and for an element of area $da$ on this surface) of internal energy, entropy and masses, respectively denoted $dU$, $dS$ and $dm_i$, and then the excess of the grand potential
\begin{gather}
d\Gamma = dU - T\,dS - \sum_i \mu_i\,dm_i, \label{grandpotential}
\end{gather}
which gives, per unit area in the present state,
\begin{gather}
\gamma = \frac{d\Gamma}{da} = U_a - T\,S_a - \sum_i \mu_i\,m_{i,a} \label{gamma}
\end{gather}
or, per unit area in some reference state (indicated by the subscript 0),
\begin{gather}
\gamma_0 = \frac{d\Gamma}{da_0} = U_{a_0} - T\,S_{a_0} - \sum_i \mu_i\,m_{i,a_0} \label{gamma0}
\end{gather}
($dU = U_a\,da = U_{a_0}\,da_0$, etc). Moreover, in \cite{Olives:2010}, we proved (\ref{surfaceinternalenergy1}) at the surface of {\it any deformable body} (in contact with a fluid), for any infinitesimal reversible thermodynamic transformation $\delta$ (from the present state to a `varied state'). As particular examples, note that such reversible transformations occur for (volume) viscoelastic solids or viscous fluids at the vanishing speed (of the material points). As a direct consequence of (\ref{defgamma0}) and (\ref{surfaceinternalenergy1}), we then obtain (23) of \cite{Olives:2010}
\begin{gather}
\delta \gamma_0 = - S_{a_0}\,\delta T  + \pi_{\rm s} \cdot \delta e_{\rm s}
- \sum_i m_{i,a_0}\,\delta\mu_i,  \label{surfacegrandpotential}
\end{gather}
which is then also valid at the surface of {\it any deformable body} (in contact with a fluid), for any infinitesimal reversible transformation $\delta$. This equation was obviously not called `Gibbs-Duhem' because it was not written by Gibbs or Duhem.

Contrary to what is said after (4) of the comment, Landau and Lifchitz \cite{Landau-Lifchitz:1967} never mentioned surface stress nor surface properties. Following the standard presentation of elasticity, they only defined the classical (volume) stress from the usual contact forces (in section 2) and after proved the classical relation between the variation of (volume) energy and the work of these (volume) stresses (in section 3).

However, surface thermodynamics is much more complex, mainly because the surface quantities are not `intuitive' (they are surface excesses on some dividing surface) and the thermodynamic variables of state of a surface are also not `intuitive' and {\it a priori} completely unknown. Note that, without a precise determination of these variables of state of the surface, any partial derivative expression has no meaning. Indeed, $\displaystyle\frac{\partial f}{\partial x_1}$ has no meaning if the other variables $x_2$, $x_3$,... of $f$ are unknown. Thus, any `definition of surface stresses' as partial derivatives of some surface energy, with respect to the components of some surface strain, {\it has no meaning if the thermodynamic variables of state of the surface are not previously perfectly identified}. But the actual problem is not to give some `definition of surface stress': it is to give the exact expression of the work of deformation of the surface, with all the variables involved in such an expression. This is not obvious and not `intuitive'. For example, the authors \cite{Nozieres-Wolf:1988,Muller-Saul:2004} recently proposed the expression of the work of deformation of the surface
\begin{gather}
\delta W_{\rm def} = (\bar{\sigma_{\rm t}} \cdot \delta \varepsilon_{\rm t} + \sigma_{\rm n} \cdot \bar{\delta \varepsilon_{\rm n}})\,da, \label{NWMS}
\end{gather}
where $\sigma$ and $\delta \varepsilon$ are the volume stress and strain tensors, the subscripts $\rm t$ and $\rm n$ respectively refer to the tangential and normal components, and the symbol $\,\bar{}\,$ indicates an excess quantity on the surface. This expression would involve six mechanical or geometrical variables of state of the surface: the three components of the tangential strain $\delta \varepsilon_{\rm t}$ and the three components of the excess of normal strain $\bar{\delta \varepsilon_{\rm n}}$. Our first aim in \cite{Olives:2010} was then to precisely determine all the variables on which the work of deformation of the surface depends and to give the exact expression of this work of deformation (without excluding any possible variable, e.g., excess of normal strain). This is why we first carefully defined the dividing surfaces and the concept of `ideal transformation' (which defines the extrapolated ideal displacements of the material points within the interface film, up to the dividing surface, and transforms the dividing surfaces into each other, making the theory consistent) in \cite{Olives:2010} section 2.1, and then presented the detailed proof of sections 2.2 and 3 of \cite{Olives:2010}. The main result of this proof is that the quantity $\delta U_{a_0} - T\,\delta S_{a_0} - \sum_i \mu_i\,\delta m_{i,a_0}$ (for an infinitesimal reversible transformation $\delta$) only depends (at the first order) on the variation of the Lagrangian surface strain tensor $e_{\rm s}$. This result enables us to write
\begin{gather}
\delta U_{a_0} - T\,\delta S_{a_0} 
- \sum_i \mu_i\,\delta m_{i,a_0} = \pi_{\rm s} \cdot \delta e_{\rm s}, \label{surfacestress}
\end{gather}
which defines the Lagrangian surface stress tensor $\pi_{\rm s}$ at equilibrium and gives the exact expression of the work of deformation of the surface: $\pi_{\rm s} \cdot \delta e_{\rm s}\,da_0$. We then defined the Eulerian surface stress tensor $\sigma_{\rm s}$, which leads to the expression of the work of deformation of the surface \cite{Olives:2010} (25)
\begin{gather}
\delta W_{\rm def} = \sigma_{\rm s} \cdot \delta \varepsilon_{\rm s}\,da = \pi_{\rm s} \cdot \delta e_{\rm s}\,da_0, \label{workdeformationsurface}
\end{gather}
(in Eulerian and Lagrangian forms). This expression differs from the above one (\ref{NWMS}) and only contains three geometrical variables, namely the components of the surface strain tensor. From (\ref{surfacegrandpotential}), we then obtained the `local' (see after \cite{Olives:2010} (21)) thermodynamic variables of state of the surface: these are the temperature $T$, the chemical potentials $\mu_i$ and the three components of the Lagrangian surface strain tensor $e_{\rm s}$. Because the variables of state are now clearly identified, we are allowed to write from (\ref{surfacegrandpotential}) the partial derivative expressions \cite{Olives:2010} (24) 
\begin{gather}
\pi_{\rm s}^{\alpha \beta} = \frac{\partial \gamma_0}{\partial e_{\rm s, \alpha \beta}}\quad \textrm {with the variables}\;(T,e_{\rm s},\mu_i).\label{surfacestress3}
\end{gather}
We insist: these results are not obvious (despite their similarity with some corresponding expressions in {\it volume} thermodynamics). Moreover, since $\delta \varepsilon_{\rm s} = \delta \varepsilon_{\rm t}$, the comparison of the two expressions (\ref{workdeformationsurface}) and (\ref{NWMS}) shows that our definition of the Eulerian surface stress $\sigma_{\rm s}$ differs from its usual definition as the excess of tangential stress $\bar{\sigma_{\rm t}}$.

The Eulerian form of (\ref{surfacegrandpotential}) is then obtained \cite{Olives:2010} (29)
\begin{gather}
\delta \gamma = - S_a\,\delta T  + (\sigma_{\rm s} - \gamma\,I) \cdot \delta \varepsilon_{\rm s}
- \sum_i m_{i,a}\,\delta\mu_i,  \label{surfacegrandpotentialeuler}
\end{gather}
which gives the corresponding partial derivative expressions
\begin{gather}
\sigma_{\rm s}^{\alpha \beta} - \gamma\,I^{\alpha \beta} = \frac{\partial \gamma}{\partial \varepsilon_{\rm s, \alpha \beta}}\quad \textrm {`with the variables}\;(T,\varepsilon_{\rm s},\mu_i)\textrm ', \label{surfacestresseuler3}
\end{gather}
with arbitrary coordinates on the surface, $I$ being the contravariant form of the metric tensor on $\rm S_{bf}$ (or identity, as a linear operator). Here, $\varepsilon_{\rm s}$ simply denotes the surface strain tensor measured from the present state (i.e., $\varepsilon_{\rm s} = 0$ in the present state, and $\varepsilon_{\rm s} = \delta \varepsilon_{\rm s}$ of (\ref{surfacegrandpotentialeuler}) in the varied state). Attention! we use quotation marks in (\ref{surfacestresseuler3}) to indicate that, contrary to $e_{\rm s}$, $\varepsilon_{\rm s}$ is not a variable of state of the surface (being taken equal to 0 in the present state, it obviously cannot characterize a state of strain in the present state): $\varepsilon_{\rm s}$ is just used to give a meaning to $\displaystyle\frac{\partial \gamma}{\partial \varepsilon_{\rm s, \alpha \beta}}$.

In the comment and \cite{Gutman:1995}, the author says that the preceding equations (\ref{surfacegrandpotentialeuler}) and (\ref{surfacestresseuler3}) show `inconsistency' (last sentence of the comment) or `are basically incorrect' (\cite{Gutman:1995} L664, line 9). This reveals a great confusion and misunderstanding of what are the Eulerian and the Lagrangian quantities (which are never distinguished nor mentioned in \cite{Gutman:1995}). In fact, (\ref{surfacegrandpotentialeuler}) and (\ref{surfacestresseuler3}) are exactly equivalent to (\ref{surfacegrandpotential}) and (\ref{surfacestress3}), the former being the Eulerian form of the (Lagrangian) latter. Indeed, with $T$ and $\mu_i$ constant, for simplicity's sake, the same equality
\begin{gather}
\delta d\Gamma = \delta W_{\rm def}
\end{gather}
may be written with the Lagrangian quantities
\begin{gather}
\delta d\Gamma = \delta (\gamma_0\,da_0) = (\delta \gamma_0)\,da_0
= \pi_{\rm s} \cdot \delta e_{\rm s}\,da_0 \label{Lagrangian}
\end{gather}
(see (\ref{gamma0}) and (\ref{workdeformationsurface})), which represents the expressions (\ref{surfacegrandpotential}) and (\ref{surfacestress3}), or with the Eulerian quantities
\begin{gather}
\delta d\Gamma = \delta (\gamma\,da) = (\delta \gamma)\,da + \gamma\,\delta da 
= (\delta \gamma)\,da + \gamma\,{\rm tr}(\delta \varepsilon_{\rm s})\,da
= \sigma_{\rm s} \cdot \delta \varepsilon_{\rm s}\,da \label{Eulerian}
\end{gather}
(see (\ref{gamma}) and (\ref{workdeformationsurface})), which represents the expressions (\ref{surfacegrandpotentialeuler}) and (\ref{surfacestresseuler3}).

Another important confusion in \cite{Gutman:1995} is to consider that the term $\gamma\,I$ in (\ref{surfacegrandpotentialeuler}) and (\ref{surfacestresseuler3}), i.e. $\gamma\,\delta da = \gamma\,{\rm tr}(\delta \varepsilon_{\rm s})\,da$ in (\ref{Eulerian}), corresponds to `forming a new surface' by `increasing the number of atoms on the surface' (\cite{Gutman:1995} L664, line 17 and L665, lines 4--5). This is incorrect. In (\ref{Eulerian}), the coefficient $\delta da = {\rm tr}(\delta \varepsilon_{\rm s})\,da$ of $\gamma$ obviously does not represent a creation of a new surface, but simply the variation of the area due to the deformation (i.e., displacement of the material points) represented by the strain $\delta \varepsilon_{\rm s}$. Note that all the terms in (\ref{Lagrangian}) and (\ref{Eulerian}) refer to the same material element of surface, the area of which is $da_0$ (fixed) in the reference state, $da$ in the present state, and $da + \delta da = da + {\rm tr}(\delta \varepsilon_{\rm s})\,da$ in the varied state.

All the assertions in \cite{Gutman:1995} are consequences of these confusions. For example, the lines 14--16 at L665 are based on the preceding confusion. Below, what is said after \cite{Gutman:1995} (8) reveals the confusion between the Eulerian variation $\displaystyle \delta \gamma = \delta(\frac{d\Gamma}{da})$ (where $da$ varies, owing to the deformation $\delta \varepsilon_{\rm s}$) and the Lagrangian one $\displaystyle \delta \gamma_0 = \delta(\frac{d\Gamma}{da_0})$ (where $da_0$ remains fixed in the reference state; from (\ref{Lagrangian}) and (\ref{Eulerian}), these two variations are related by $(\delta \gamma_0)\,da_0 = (\delta \gamma)\,da + \gamma\,{\rm tr}(\delta \varepsilon_{\rm s})\,da$).

Let us now introduce the {\it creation} of a new surface which, with our notations, is represented by the variation of area $\delta da_0$ in the reference state. The corresponding variation of area $\delta da$, between the present state and the varied state, will then be the sum of two terms
\begin{gather}
\delta da = \delta da_{\rm def} + \delta da_{\rm cr}
= {\rm tr}(\delta \varepsilon_{\rm s})\,da + (\frac{da}{da_0})\,\delta da_0, \label{deformation-creation}
\end{gather}
the first one being due to the deformation $\delta \varepsilon_{\rm s}$ (variation of $da$ when $da_0$ remains fixed), and the second one to the creation $\delta da_0$ (variation of $da$ when the present state of strain remains fixed, i.e. when $\delta \varepsilon_{\rm s} = 0$). Our Lagrangian equation \cite{Olives:2010} (28)
\begin{gather}
\delta dU = T\,\delta dS + \pi_{\rm s} \cdot \delta e_{\rm s}\,da_0
+ \gamma_0\,\delta da_0 + \sum_i \mu_i\,\delta dm_i, \label{surfaceinternalenergy3}
\end{gather}
which clearly shows the deformation term $\pi_{\rm s} \cdot \delta e_{\rm s}\,da_0$ and the creation term $\gamma_0\,\delta da_0$, may then be written in the Eulerian form
\begin{gather}
\delta dU = T\,\delta dS + \sigma_{\rm s} \cdot \delta \varepsilon_{\rm s}\,da
+ \gamma\,\delta da_{\rm cr} + \sum_i \mu_i\,\delta dm_i, \label{surfaceinternalenergyeuler3}
\end{gather}
with the corresponding deformation term $\sigma_{\rm s} \cdot \delta \varepsilon_{\rm s}\,da$ and 
creation term $\gamma\,\delta da_{\rm cr}$ ($\gamma_0\,\delta da_0 = \gamma\,\delta da_{\rm cr}$, since $\displaystyle \frac{\delta da_{\rm cr}}{\delta da_0} = \frac{da}{da_0}$), i.e.
\begin{gather}
\delta dU = T\,\delta dS + (\sigma_{\rm s} - \gamma\,I) \cdot \delta \varepsilon_{\rm s}\,da
+ \gamma\,\delta da + \sum_i \mu_i\,\delta dm_i \label{surfaceinternalenergyeuler4}
\end{gather}
(since $\delta da_{\rm cr} = \delta da - {\rm tr}(\delta \varepsilon_{\rm s})\,da$, from (\ref{deformation-creation}); compare the above equation with \cite{Olives:2010} (1) for a fluid--fluid surface, where $\sigma_{\rm s} = \gamma\,I$).

\section*{Acknowledgments}

\noindent We acknowledge financial support from CINaM-CNRS and ANR-08-NANO-036.


\begin{thebibliography}{5}
\bibitem{Olives:2010} Olives J 2010 {\it J. Phys: Condens. Matter} {\bf 22} 085005
\bibitem{Gutman:1995} Gutman E M 1995 {\it J. Phys: Condens. Matter} {\bf 7} L663--7
\bibitem{Landau-Lifchitz:1967} Landau L and Lifchitz E 1967 {\it Th\'eorie de l'\'elasticit\'e} (Moscow: Mir)
\bibitem{Nozieres-Wolf:1988} Nozi\`eres P and Wolf D E 1988 {\it Z. Phys.} B {\bf 70} 399--407
\bibitem{Muller-Saul:2004} M\"uller P and Sa\'ul A 2004 {\it Surf. Sci. Rep.} {\bf 54} 157--258
\end{thebibliography}
\end{document}